\documentclass[prx,twocolumn,superscriptaddress]{revtex4-1}

\usepackage{color}
\usepackage[english]{babel}
\usepackage[T1]{fontenc}
\usepackage[latin9]{inputenc}
\usepackage{latexsym}
\usepackage{float}
\usepackage{amsmath}
\usepackage{amsfonts}
\usepackage{amssymb}
\usepackage{graphicx}
\usepackage[export]{adjustbox}
\usepackage{times}   
\usepackage{esint}
\usepackage[unicode=true,pdfusetitle,
 bookmarks=false,colorlinks=true,citecolor=blue,urlcolor=blue,linkcolor=red]{hyperref}
\newcommand*{\nolink}[1]{\begin{NoHyper}#1\end{NoHyper}}

\makeatletter

\@ifundefined{textcolor}{}
{%
 \definecolor{BLACK}{gray}{0}
 \definecolor{WHITE}{gray}{1}
 \definecolor{RED}{rgb}{1,0,0}
 \definecolor{GREEN}{rgb}{0,1,0}
 \definecolor{BLUE}{rgb}{0,0,1}
 \definecolor{CYAN}{cmyk}{1,0,0,0}
 \definecolor{MAGENTA}{cmyk}{0,1,0,0}
 \definecolor{YELLOW}{cmyk}{0,0,1,0}
}

\@ifundefined{date}{}{\date{}}
\AtBeginDocument{
  
}
\makeatother

\newcommand{\SAVE}[1]{}

\newcommand{\rmn}[1]{{{#1}}}

\newcommand{\sectionn}[1]{{\textit{\underline{#1:}}}}

\newcommand{\ncnf}{NaCaNi$_2$F$_7$}

\begin{document}
\renewcommand\abstractname{}

\title{Dynamical structure factor of the three-dimensional quantum spin liquid candidate NaCaNi$_2$F$_7$} 
\author{Shu Zhang}
\affiliation{Department of Physics and Astronomy, Johns Hopkins University, Baltimore, MD 21218} 
\affiliation{Institute for Quantum Matter, Johns Hopkins University, Baltimore, MD 21218}
\author{Hitesh J. Changlani}
\affiliation{Department of Physics, Florida State University, Tallahassee, Florida 32306}
\affiliation{Department of Physics and Astronomy, Johns Hopkins University, Baltimore, MD 21218} 
\affiliation{Institute for Quantum Matter, Johns Hopkins University, Baltimore, MD 21218}
\author{Kemp W. Plumb}
\affiliation{Department of Physics, Brown University, Providence, RI 02912}
\author{Oleg Tchernyshyov}
\affiliation{Department of Physics and Astronomy, Johns Hopkins University, Baltimore, MD 21218} 
\affiliation{Institute for Quantum Matter, Johns Hopkins University, Baltimore, MD 21218}
\author{Roderich Moessner}
\affiliation{Max-Planck Institute for the Physics of Complex Systems, 01187 Dresden, Germany}
\date{\today}

\begin{abstract}
We study the spin-1 pyrochlore material NaCaNi$_2$F$_7$ with a combination of molecular dynamics simulations,
stochastic dynamical theory and linear spin wave theory.
The dynamical structure factor from inelastic neutron scattering is 
well described with a near-ideal Heisenberg Hamiltonian incorporating small anisotropic terms 
{and weak second-neighbor interactions}. 
We find that all three approaches reproduce remarkably well the momentum dependence of the 
{scattering intensity as well as its energy dependence with the exception of the lowest energies.}
These results are notable in that (i) 
{the data show a complete lack of sharp}
quasiparticle excitations in momentum space over much, if not all, of the energy range; 
(ii) linear spin-wave theory appears to apply in a regime 
where it would be expected to fail for a number of reasons. 
We elucidate what underpins these surprises, 
and note that basic questions about the nature of quantum
  spin liquidity in such systems pose themselves as a result.
\end{abstract}

\maketitle
\sectionn{Introduction} 
Quantum spin liquids~\cite{AndersonRVB} are enigmatic phases 
of matter characterized by the absence of symmetry breaking and conventional quasiparticles (magnons). 
The search for their realisation in actual magnetic materials has targeted, but  is
 not limited to, materials involving the geometrically 
 frustrated~\cite{FrustratedMagnetism} triangular, kagome~\cite{KanodaNgQSL} and pyrochlore~\cite{GGGPyrochlore} geometries with low spins. 
 
Indeed, while there have been significant efforts to synthesize quantum spin liquid materials in spin-1/2 systems in two dimensions, fewer efforts 
have been devoted to
three dimensions, see Ref.~\cite{KanodaNgQSL} for a review. 
This strategic choice is not without reason: high lattice coordination number and high spin typically 
suppress quantum fluctuations, and thus in particular
favor conventional forms of magnetic order over quantum spin liquids in three dimensions. However, 
it is now clear that this perspective is too pessimistic: as a matter of principle, we know that certain
types of spin liquid -- in particular, Coulombic U(1) spin liquids -- can exist in $d=3$ but not in $d=2$~\cite{MS3dRVB,HermeleP}; 
and as a matter
of practice, it looks as if quantum spin liquid phases need 
by no means be restricted to $S=1/2$ exclusively~\cite{IqbalThomale2018}. 

Despite several recent advances in the field, however,  our understanding  of the actual properties of 
low-spin Heisenberg spin liquids in three dimensions is very limited, as they 
are beyond the scope of practically all exact or controlled approximate theoretical schemes. 
We are at a loss to describe either their ground states or excitation spectra, unlike Ising models like spin ice, 
where the simplest quantum versions~\cite{MS3dRVB,HermeleP} are 
amenable to {quantum Monte Carlo (QMC) simulations}~\cite{IsakovP,ShannonL}. Experimental data is therefore
a particularly  indispensable guide for our understanding of these magnets, for an early review see~\cite{ZinkinHarris}.

In the quest to identify quantum spin liquids in real materials~\cite{KnolleMonssner2018}, one relies heavily on characteristic signatures in the magnetic excitation spectra, as their ground states are often largely featureless. By contrast, 
the excitations of spin liquids can be downright spectactular, including in particular fractionalised~\cite{RajaramanReview} and
other unusual emergent quasiparticles such as spinons in the spin-1/2 Heisenberg antiferromagnet chain~\cite{Bethe1931,LakePRL2013,Mourigal2013}, Majorana fermions in the Kitaev honeycomb model~\cite{KitaevH,Banerjee2017,Matsuda2018},  and magnetic monopoles~\cite{CMS2008} and photons in the U$(1)$ spin liquid~\cite{MS3dRVB,HermeleP,ShannonL}. 

The dual challenge is thus to identify novel behaviour in experimental data on candidates quantum spin liquid
materials in $d=3$, and to devise a theoretical framework for understanding the underlying behaviour. 
Here, we report progress for the fluoride pyrochlore \ncnf{}~\cite{KrizanCava2015NCNF},  a prime $S=1$ 
quantum spin liquid candidate.

\begin{figure}[t]
\includegraphics[width=0.5\linewidth]{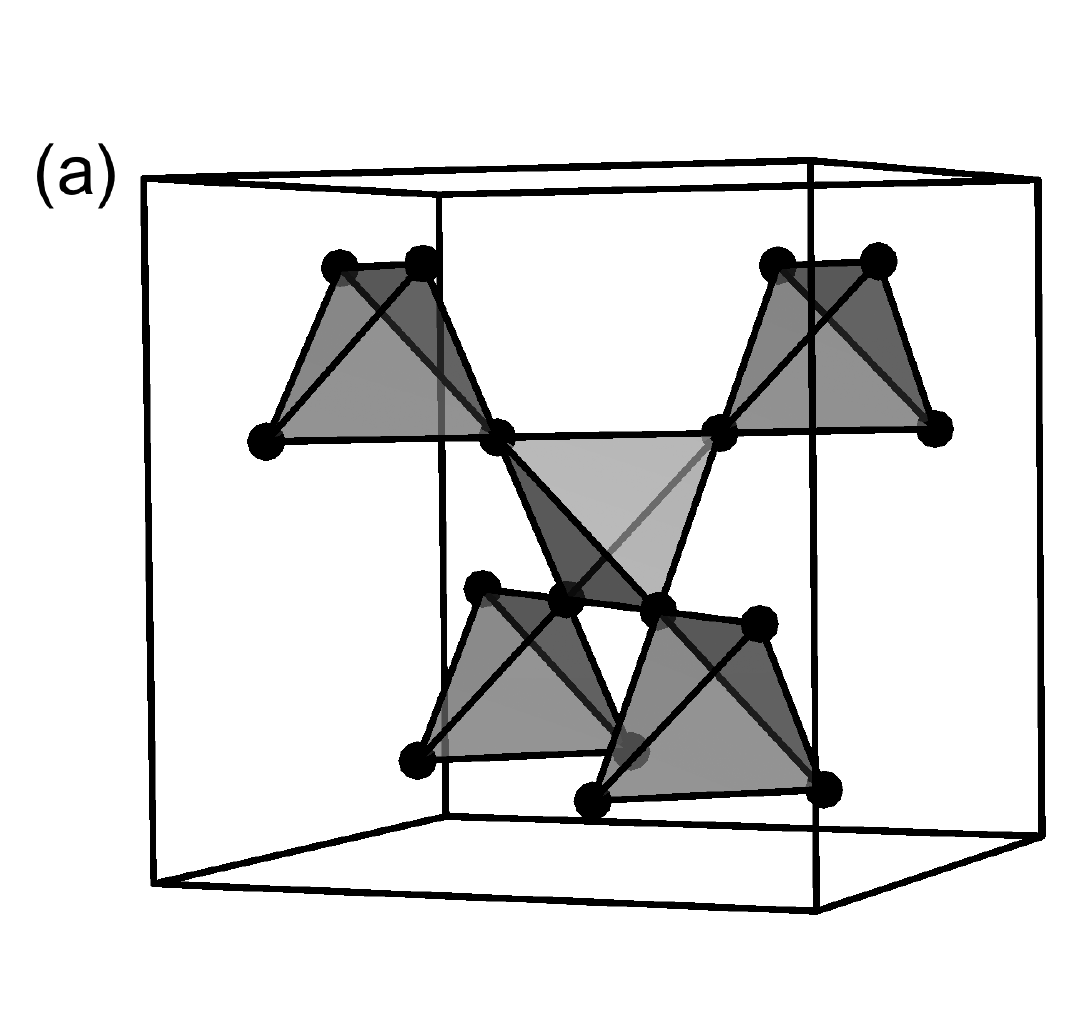}
\hfill
\includegraphics[width=0.45\linewidth]{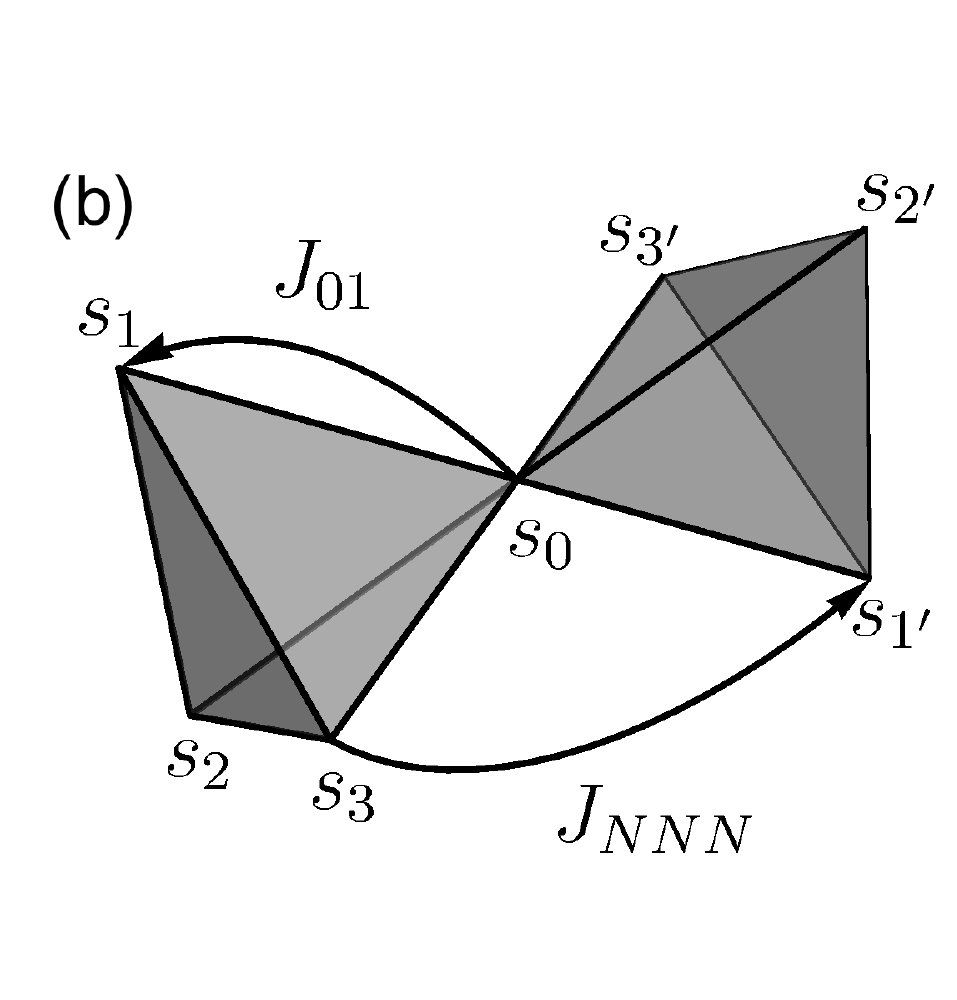}
\caption{(a). Pyrochlore lattice in one cubic unit cell. (b). Nearest-neighbor and next-nearest-neighbor interactions.}
\label{fig:interaction} 
\end{figure}

In \ncnf{}, the magnetic Ni$^{2+}$  ions reside on the three-dimensional pyrochlore lattice (Fig.~\ref{fig:interaction}),
where what little is known theoretically about quantum Heisenberg models for $S=1/2$ and $S=1$ 
points towards 
quantum spin liquid behavior~\cite{CanalsLacroix1998,IqbalThomale2018},
while the classical case is well-established to be a Coulomb spin liquid~\cite{Villain1979, MoessnerChalker1998PRL, MoessnerChalker1998PRB}, whose
many-body dynamics is by now fairly well understood~\cite{MoessnerChalker1998PRL, MoessnerChalker1998PRB, ConlonChalker2009}.

We analyze the magnetic excitation spectrum
obtained by inelastic neutron scattering on \ncnf{} in Ref.~\cite{PlumbBroholm2017}, which we supplement
with new data from a different experiment.
We present three tractable complementary theoretical approaches that reproduce 
the dynamical structure factor $\mathcal{S}(\mathbf q,\omega)$ 
for all momenta $\mathbf q$ and for
a broad range of energies $\omega$. At the highest energies, the quality of the agreement differs between models (and becomes harder to assess on account of a considerable phonon background). At low energies, we find the well-known pinch-point motifs, while at intermediate energies, characteristic structures complementary to the pinch points appear~\cite{Yan2018,Mizoguchi2018}. Overall, the main disagreement between experiment and theory appears 
at the lowest energies, as discussed below. 

In the light of the abovementioned challenges posed by three-dimensional quantum spin liquids, 
the capacity of our relatively simple approaches to yield a wide-ranging account of the observed
dynamics is as striking as it is encouraging for the study of other yet unexplored systems and models
in this class. We therefore include a discussion of the broader implications of our results about the nature
of the quantum dynamics in such a setting, which we believe may be of importance well beyond
the material studied here. 

From a standpoint of basic phenomenology, we particularly emphasize that  none of 
the employed theoretical approaches relies on the existence
of branches of spin waves or other quasiparticles with well-defined wavevector $\mathbf q$ and frequency $\omega$, 
nor do they require the presence of delicate quantum coherence. At the same time, {\it linear spin wave theory}
is among the methods which is successful in this context! 

\begin{figure*}[t]

\includegraphics[width=0.48\linewidth]{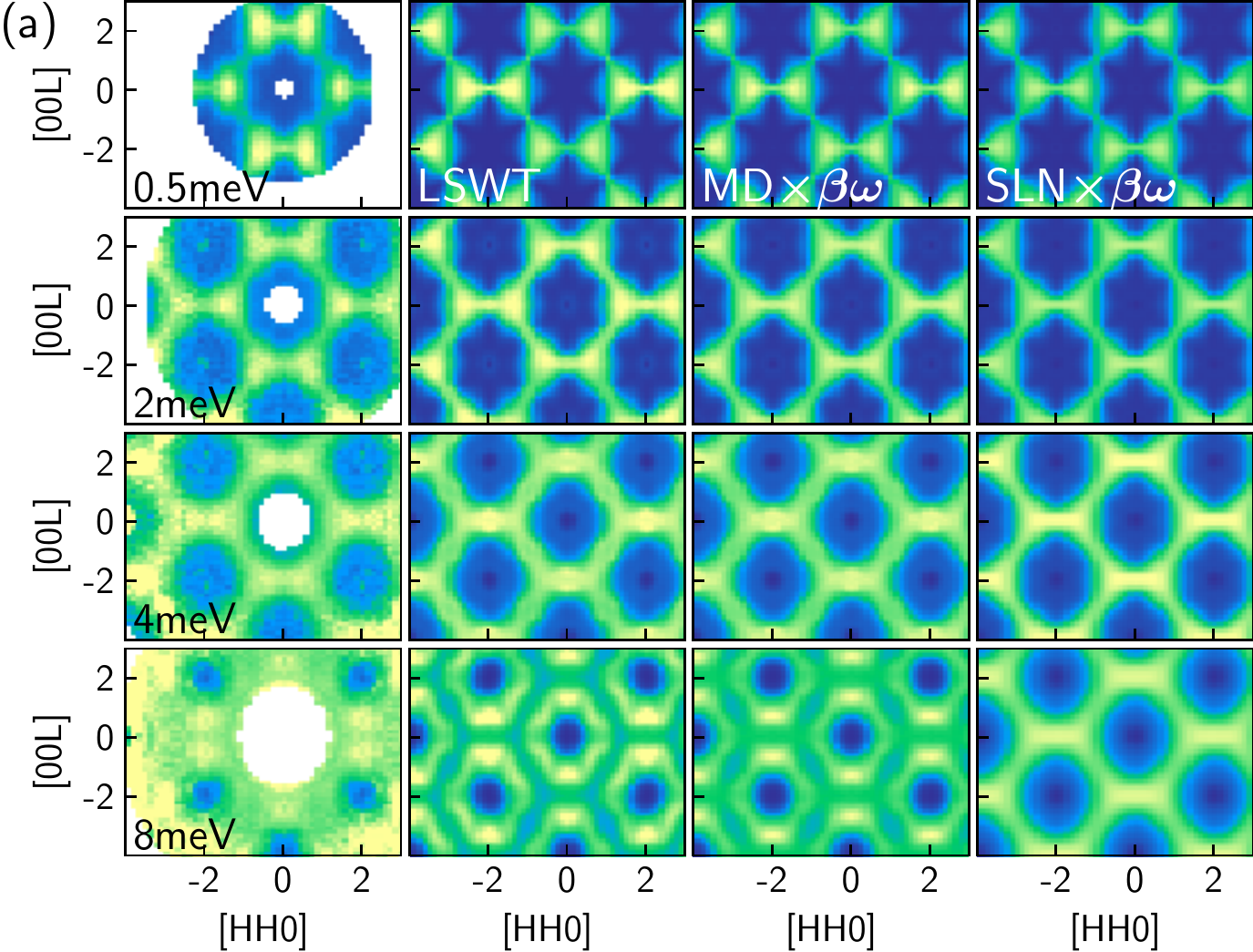}
\hfill
\includegraphics[width=0.48\linewidth]{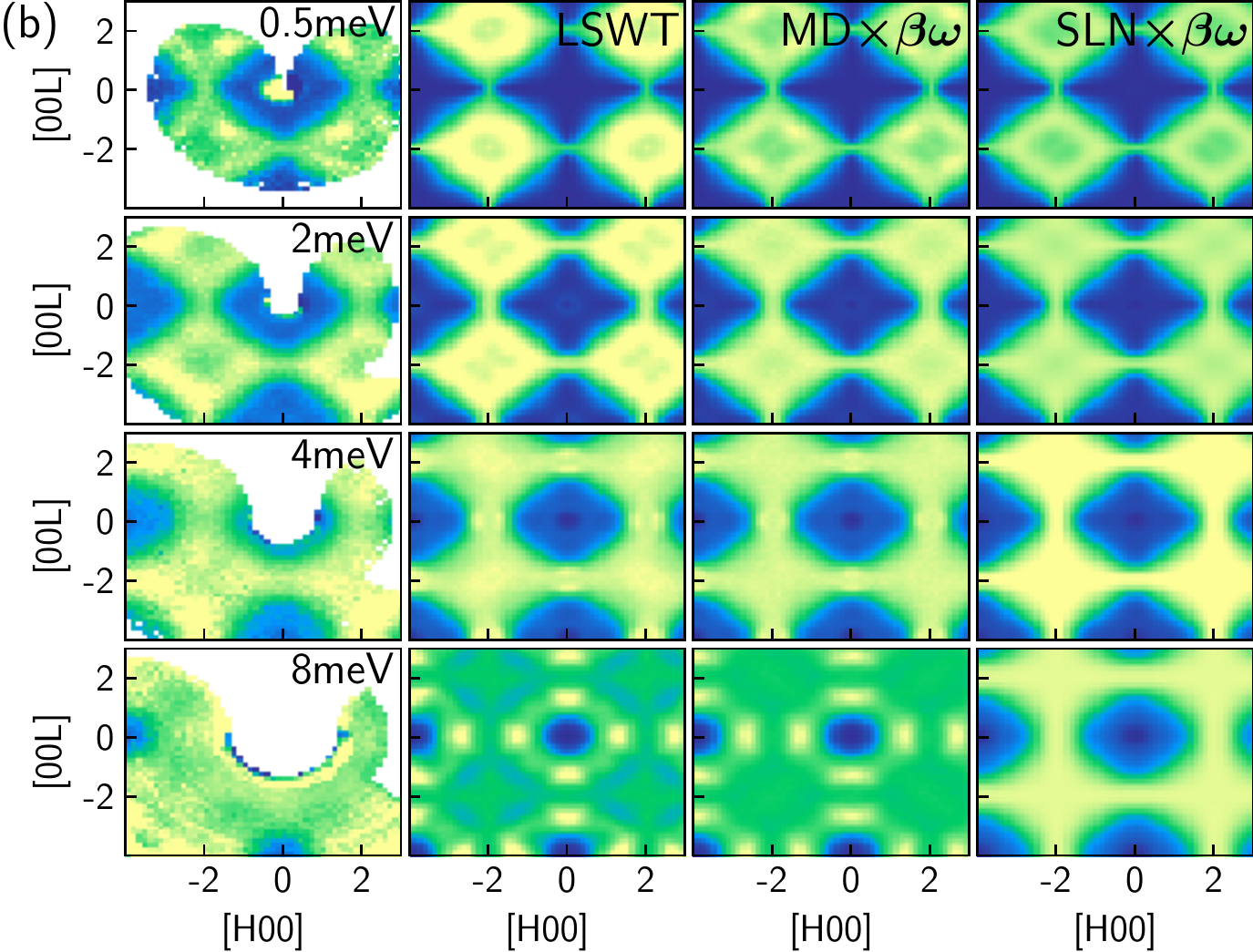}
\includegraphics[width=0.95\linewidth, right]{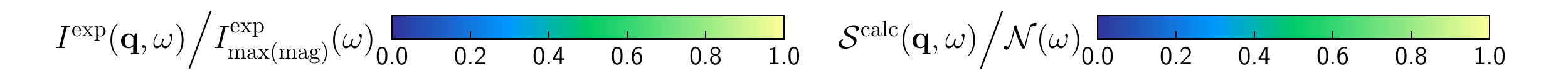}
\caption{(Color)
Momentum dependent dynamical structure factor in [HHL] plane (a) and [H0L] plane (b) at constant energies: inelastic neutron experiment compared to linear spin wave theory, molecular dynamics, and the stochastic model. The data in panel (a) were collected on CNCS and the data in panel (b) were collected on MACS. Raw neutron scattering intensity has been corrected by the magnetic form-factor for Ni$^{2+}$~\cite{Brown:06}.   To focus on the wavevector dependence, the data are rescaled for each value of energy, for the experimental data, by the  maximum magnetic scattering intensity; and by the maximum intensity in the MD simulations for the six theory panels, with an additional factor $ \beta \omega$ between MD/SLN and LSWT (Eq.~(\ref{eq:correspondence}), see text), ${\mathcal N} (\omega) = \beta \omega {\mathcal S}^\textrm{MD}_\textrm{max}(\omega)$, where $\beta = 1/ k_B T$.
}
\label{fig:patterns} 
\end{figure*}

\begin{figure}[h!]
\includegraphics[width=\linewidth]{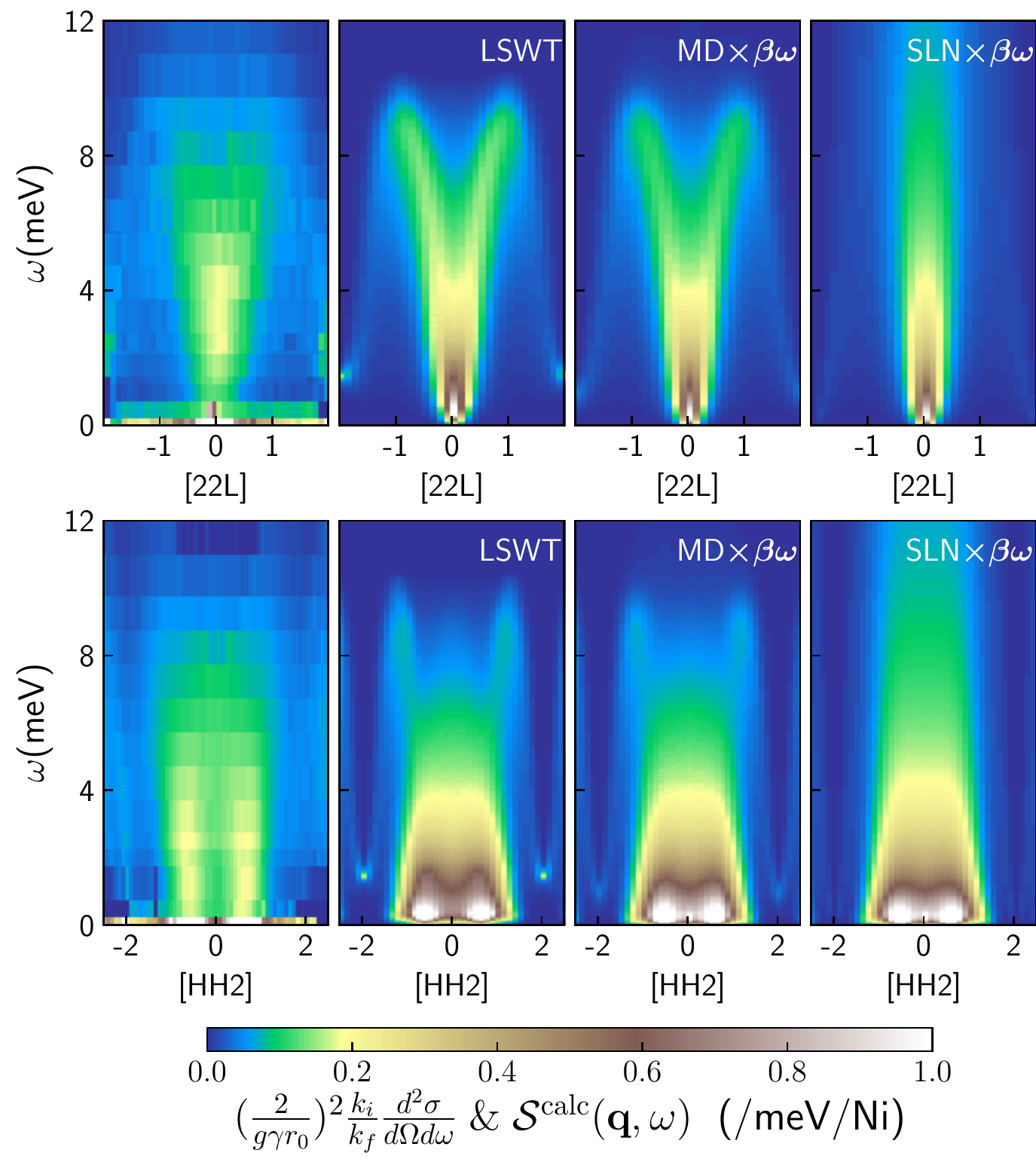}
\caption{(Color)
Energy dependence of dynamical structure factor along momentum cuts [22L] and [HH2]. Neutron scattering intensity is in absolute units (see Appendix~\ref{Appendix:experiment}). \rmn{Rescaled MD and LSWT in particular reproduce well the shape of the broad dispersive curve, 
disagreeing mainly at the lowest energies, while SLN fails to capture high-energy structure.}
}
\label{fig:dispersion} 
\end{figure}

\begin{figure}[t]
\includegraphics[width=\linewidth]{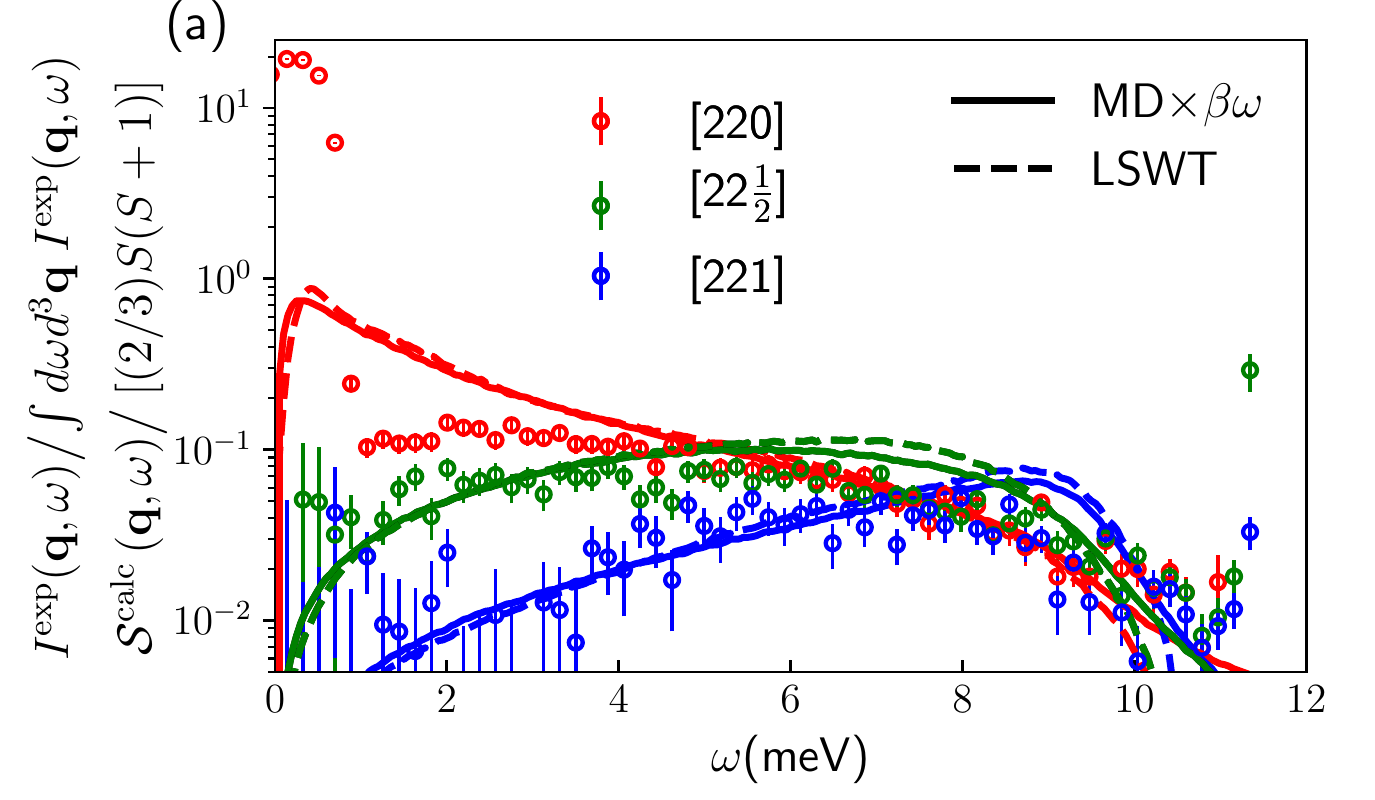}
\includegraphics[width=\linewidth]{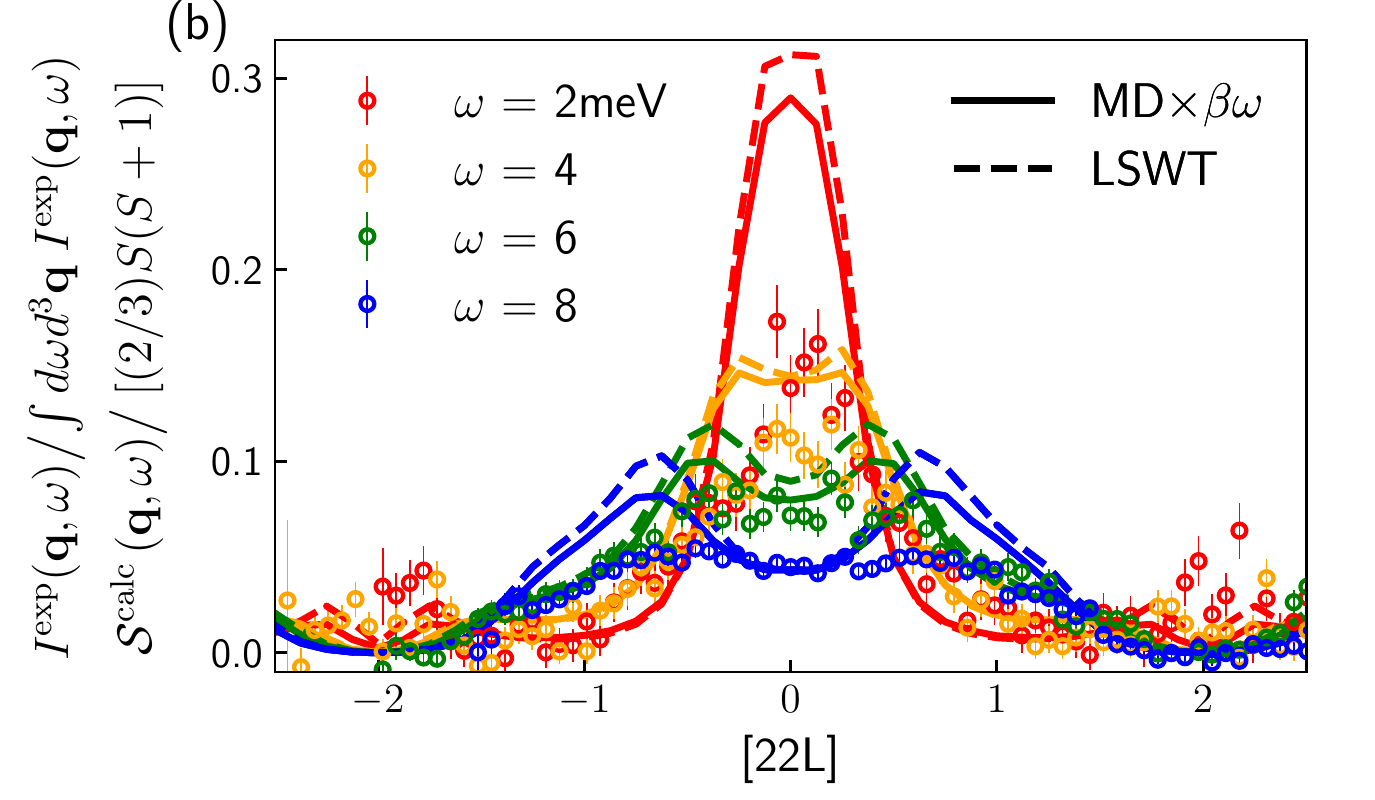}
\caption{(Color)
Comparison of dynamical structure factor between experiment and MD/LSWT. (a) Energy-dependence at $\mathbf{q} = $ [220], [22$\frac{1}{2}$] and [221]. Log scale is used for the y-axis to include the quasi-elastic signals. (b) Momentum dependence along [22L] at $\omega = $ 2,4,6 and 8meV. The neutron scattering intensity $I^\mathrm{exp} (\mathbf{q},\omega)$ is background subtracted and normalized by the total spectral weight $\int d\omega d^3 {\mathbf q} \, I^\mathrm{exp} (\mathbf{q},\omega)$. MD (solid lines) and LSWT (dashed lines) data $\mathcal{S}^\mathrm{calc}(\mathbf{q},\omega)$ are normalized by $(2/3)S(S+1)$, under an isotropic approximation to the sum rule.}
\label{fig:methods-comparison} 
\end{figure}

We next introduce our model, and the main results on the dynamical structure factor of \ncnf are presented in the form of a comparison of scattering intensities as a function of momentum (Fig.~\ref{fig:patterns}) and energy (Figs.~\ref{fig:dispersion}, \ref{fig:methods-comparison}). Besides their interpretation and discussion,  for the 
methodologically interested reader, we collate all necessary technical information in a set of self-contained technical
appendices.

\sectionn{Model and methods} 
We use the Hamiltonian (Fig.~\ref{fig:interaction})
\begin{equation}
H \!=\!\frac{1}{2}\sum_{ij}\sum_{\mu\nu}J^{\mu\nu}_{ij} s^{\mu}_i s^{\nu}_j,
\end{equation}
where subscripts $i$ and $j$ refer to lattice sites and superscripts $\mu$ and $\nu$ refer to Cartesian components of spins in the global frame.
The interaction matrix is parameterized by four exchange parameters between nearest neighbors $J_{01} = (J_2,J_4,J_4;-J_4,J_1,J_3;-J_4,J_3,J_1)$ with
$J_{1}\!=\!J_{2}\!=\!3.2(1)$~meV, $J_{3}\!=\!0.019(3)$~meV,
$J_{4}\!=\!-0.070(4)$~meV, and isotropic between next-nearest-neighbors $J_{NNN}\!=\!  -0.025(5)$~meV, obtained by fitting to the equal-time 
correlations by some of us in a previous study~\cite{PlumbBroholm2017}. The interaction matrices for other pairs follow
from appropriate symmetry transformations.

The methods utilized are, firstly, molecular dynamics (MD) simulations of the pyrochlore
magnet~\cite{MoessnerChalker1998PRL} where the semiclassical Landau-Lifshitz equations of motion for the
spins are integrated numerically, averaged over initial conditions obtained from Monte Carlo simulations
of our Hamiltonian at the temperature $T \!=\! 1.8$~K. Secondly, we employ a {self-consistent Gaussian approximation 
adapted to frustrated magnets}~\cite{GaraninCanals} and extended into a stochastic model for their dynamics by~\nolink{\citeauthor{ConlonChalker2009}}~\cite{ConlonChalker2009}, which we refer to as stochastic large-$N$ (SLN). 
Thirdly, we use the linear spin-wave theory (LSWT) to describe spin dynamics near 
a low-energy state (again averaged over an ensemble  obtained form Monte Carlo simulations). Details of these are provided in the 
appendix.

The central object of investigation are the dynamical spin correlations as captured by the structure factor in energy-momentum space
\begin{equation}
\begin{aligned}
&{\mathcal S}({\mathbf q}, \omega) = \sum_{\mu \nu} (\delta_{\mu\nu} - \frac{q_\mu q_\nu}{q^2}) \\
& \qquad \times  \frac{1}{2\pi N} \sum_{i,j = 1}^{N} \int_{-\infty}^\infty dt \,e^{-i \mathbf q \cdot(\mathbf r_i - \mathbf r_j) + i \omega t}  \langle s_i^\mu (t) s_j^\nu (0)\rangle.
\label{eq:dsf}
\end{aligned}
\end{equation}
The classical expression is given above and the quantum expression is sensitive to time order of the spin operators (see Appendix~\ref{Appendix:quantum}). 
Throughout this work,  we rescale the two "classical" approaches MD and SLN by {$\beta\omega$} to make comparison with LSWT. The factor essentially arises from the classical 
equipartition-based suppression of the low-$T$ intensity, 
as opposed to the non-vanishing matrix elements in the quantum case, see Eq.~(\ref{eq:correspondence}). 

\sectionn{Results}
The dynamical structure factor obtained 
experimentally and by  the three theories 
is depicted   as a function of wavevector 
in a set of cuts at various energies,  Fig.~\ref{fig:patterns},  and  as a function of energy
along a set of paths through reciprocal space, Fig.~\ref{fig:dispersion}.

Fig.~\ref{fig:patterns} displays normalized momentum cuts  in the [HHL] and [H0L] planes 
at energies 0.5, 2, 4 and 8 meV. 
At low energies, the pinch points characteristic of Heisenberg pyrochlore magnets are 
clearly visible.
This is in itself interesting, as {the presence of} well-defined pinch points implies that each tetrahedron has vanishing total
magnetization~\cite{IsakovPRL2004,HenleyCoulomb}. In general, however, adjacent tetrahedra cannot both
be in spin singlet states, as their total spin operators do not commute. 
Therefore, while for the classical theories, the pinch points sharpen as $\sqrt{T}$ as $T$ is lowered~\cite{ZinkinT},
for $S=1/2$ they were found to be quite smeared out~\cite{CanalsLacroix1998},
becoming sharper as $S$ increases. For $S=1$, a theoretical prediction for the full-width at half maximum of the pinch point in the static correlations at [002] (located at $(0,0,4\pi)$ in reciprocal space) of 
$\delta q^{\mathrm{FWHM}}_{\mathrm{PP}}=4\pi/3$~\cite{IqbalThomale2018} is comparable to the value $\approx \pi$  extracted form the low-$T$ experimental data.

As the energy increases, the overall intensity distribution changes little initially, but whatever sharp features were present
wash out{; e.g.,} the intensity minimum in the scattering rhombus around [202] is slowly filled in and the pinch points broaden. 
At higher energies, the experimental signal is increasingly polluted by the phonon background at large $q$, but it is still possible to 
identify a qualitative rearrangement of the weight, especially in the [HHL] data,  
with the area around the pinch-points growing into prominent  
pairs of "half-moons" features at 8 meV. These have 
recently been identified  as a dispersing complement to the pinch points~\cite{Yan2018,Mizoguchi2018}. This feature is present in MD and LSWT, 
but not in SLN, which is relaxational and
does not capture the spin precession
at high frequencies. 

We next turn to the energy dependence of the data, depicted in Fig.~\ref{fig:dispersion}, with additional
cuts from a different neutron instrument (see Appendix~\ref{Appendix:experiment}) presented in Fig.~\ref{fig:methods-comparison}.
The general shapes of experiment and MD/LSWT are very similar---a broad signal with a vertical
appearance reminiscent of a fountain. SLN fails to capture the high-energy structure, which can therefore
be ascribed to the precessional spin dynamics not captured by this method; otherwise, the theory plots essentially
agree with one another.

The largest disagreement between theory and experiments occurs at low frequencies, especially around [220], 
where a large increase of the 
experimental signal below $\omega=1$~meV is not reflected in theory, Fig.~\ref{fig:methods-comparison}. We return
to the issue of the low-frequency regime in the discussion below.

\sectionn{Discussion} We next address several more  
general questions arising from our central observation of what we believe is remarkable agreement
between theory and experiment at all but the lowest energies on one hand, and between MD and LSWT
on the other. We start with the latter, which is quite unexpected:
the $S=1$ Heisenberg pyrochlore antiferromagnet
 unifies several reasons why LSWT should break down. Instead, it 
works (un)reasonably well, as  evidenced in the comparison by eye with experiment as well as
in the detailed quantitative agreement with MD, Fig.~\ref{fig:methods-comparison}.
The inauspicious ingredients are, firstly, the absence of a state with long-range order around which to perturb,
the existence of which would have 
 guaranteed a Goldstone mode as long-lived magnon
   excitation. Other settings which lack long-range order, such as the $S=1/2$ or $S=1$ Heisenberg
   chain, instead show a breakdown of LSWT, as their respective low-energy descriptions involve 
   not the gapless magnons but rather fractionalized $S=1/2$ spinons
   and Haldane's famous gap.  Secondly, the spin length, $S=1$, really is
   not particularly large in our setting, so that one would generically 
   expect at least considerable quantum renormalization effects, 
   all the more so since the classical local exchange field, a central feature in suppressing fluctuations, 
   is reduced as a result of geometric frustration from $6S$ in a ferromagnet to  $2S$.
   Finally, a finite fraction of the spin-wave modes live at or near zero energy in LSWT, which implies the onset of the many-particle continuum {\it already at the bottom of the single-particle spectrum}. Above this onset, spin waves are generally
  expected to cease to be a useful description of the excitation spectrum~\cite{ChernyshevPRB2009}.

So why does linear spin wave theory nonetheless work so well? LSWT actually finds another route to work---it is not a theory of universal low-energy hydrodynamic excitations, but more a description of the statistically
typical short-to-intermediate time behavior, which in fact does not do a good job precisely at the lowest energies; thus in the end conforming to at least a subset of the above expectations. 

\begin{figure}
\includegraphics[width=0.9\linewidth]{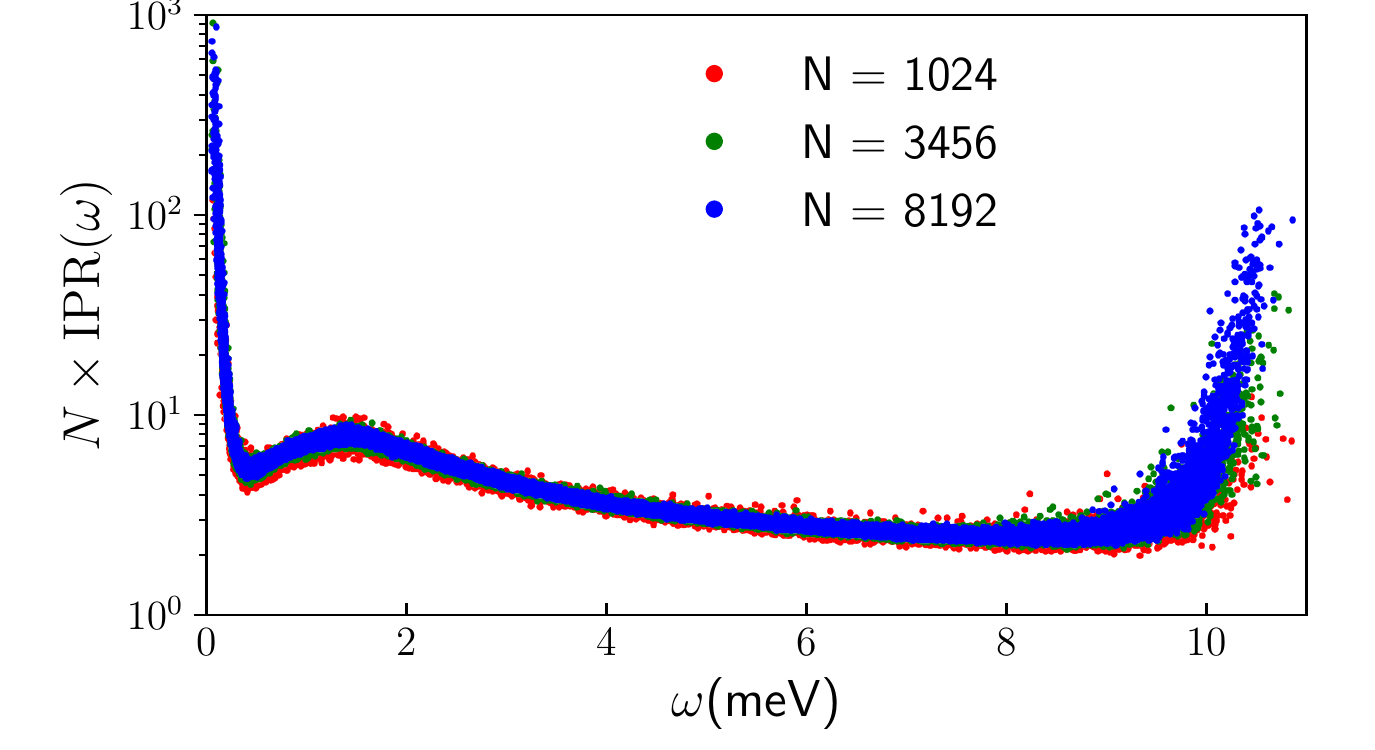},
\caption{(Color)
Inverse participation ratio (IPR) of the normalized real-space spin wave modes  $\psi( \mathbf{r}, \omega)$ as a function of frequency for different system sizes. 
See Appendix~\ref{Appendix:rswt} for the calculation details. The modes
are delocalized everywhere except for possibly at the edges of the spectrum.}
\label{fig:ipr} 
\end{figure}

To see this, think of the (near-)zero 
frequency modes responsible for motion between (near-)degenerate ground states, and oscillatory 
excitations around these as driving this motion~\cite{MoessnerChalker1998PRL, MoessnerChalker1998PRB, ConlonChalker2009}. The latter have finite frequency and finite scattering rates. The central ingredient is
that, statistically, there appears to be no difference 
between the fast spectra of states visited as the slow modes evolve, so that such motion is not reflected 
in the broad spectra we consider here. One may expect sample-to-sample fluctuations due to the disorder inherent in the randomly sampled starting configuration to be small, not least because 
the almost uniform exchange field implies that  disorder is mainly in off-diagonal,
terms of the dynamical matrix. In keeping with this, we find self-averaging in practice as only a few configurations are needed to obtain smooth spectra for large system sizes (see Appendix~\ref{Appendix:rswt}); and as expected for weak disorder in three dimensions, 
the spin wave modes away from the band edges are delocalized, as  diagnosed by the scaling of their inverse participation ratio with system 
size (Fig.~\ref{fig:ipr}). 

This also resolves the conundrum why scattering of the spin waves---unavoidable
as the slow modes evolve~\cite{MoessnerChalker1998PRB}---does not invalidate the spin wave picture. 
Given LSWT  finds a broad
continuum in frequency space to begin with, any further broadening of an individual mode 
due to its limited lifetime will be small in temperature $T$, and therefore
parametrically smaller than the total (largely $T$-independent) bandwidth. 
Therefore, unlike in the case of an initially sharp mode, lifetime broadening 
is insignificant. 

Regarding the low-$T$ limit, the zero modes mentioned above have no dynamics in LSWT {(that is just the 
statement that their frequency is zero)}. The motion along the ground-state manifold is thus essentially
frozen out, and LSWT in fact fails completely to capture their motion arising from scattering high-energy
excitations, which is present in (not rescaled) MD and SLN theories.

From the preceding paragraph, it is clear that our comparison is
not particularly sensitive to the detailed nature of the 
low-frequency behavior. Indeed, it has been a common theme of several recent studies
of exotic magnetic dynamics that  scattering away from low energies are most instructive. 
While this part of the spectrum is not universal, it may permit simple models, e.g.\ in terms of deconfined spinons
in the case of the Heisenberg chain~\cite{LakePRL2013} which in its detailed agreement with experiment may be more convincing than
the relatively featureless, and fragile, low-energy universal features{.This is all the more so since, like here, this portion
 is often experimentally harder to access.  Furthermore, 
the most characteristic aspects of the spin liquid ground states are topological in nature and as such
invisible to {experimental probes that couple to local correlations} anyway~\cite{KnolleMonssner2018}.}

\rmn{Further complicating the low-energy analysis is the presence of disorder and (partial) freezing~\cite{KrizanCava2015NCNF,CaiLuke2018}, which will need to be included in a separate
nontrivial modeling effort~\cite{ChalkerD}. 
Also, further small terms in the Hamiltonian to which the previous fitting procedure may
be insensitive, can additionally lead to shifts of weight on a scale which is small in absolute units but nonetheless 
notable at low energies. Further, to accurately model a low-energy window comparable in size to the temperature, 
a more detailed correspondence
between classical and quantum calculations than our simple rescaling ansatz would be needed.}  

While bearing all of this in mind, we emphasize the complete 
absence of sharp quasiparticle peaks characteristic of magnons with well-defined momenta and energies
both in theory and experiment. 
This reflects the spatially disordered nature of the spin configurations in our classical theory, while posing the question
about the appropriate description of the corresponding low-temperature quantum state. In particular, it will be interesting
to know if the small-spin pyrochlore Heisenberg antiferromagnet exhibits no well-defined quasiparticle excitations at all.

The final basic issue raised by our study is the role of the "quantumness" in this compound. The relative
 success of fully classical
modeling 
across a broad range of energies, at temperatures far below the Curie-Weiss scale, is rather unexpected. 
The low-energy discrepancies discussed above seem like a small price to pay for the 
huge simplicity of the theoretical approaches we have employed. This calls for an experiment on analogous compounds with
larger spin, to investigate whether the low-energy regime will be better modeled while retaining the other features
already successfully accounted for. 

Employing semi-classical modelling for what "ought to be" a quantum spin liquid is not without precedent. 
This was done for the Kitaev honeycomb model~\cite{KitaevH}, which has the benefit that the availability of 
an exact solution of the dynamical structure factor~\cite{KnollePRL2014} of the spin liquid allows for  a reliable comparison in detail. 
There~\cite{SamarakoonBatista2017}, 
the high-frequency portion of the response was accounted for modulo a reasonable  amount of data post-processing,
while the physics related to the emergent fluxes at low energies---the most direct manifestation of fractionalization---remained
inaccessible. 

This of course suggests a similar scenario here, namely that qualitative signatures of a quantum spin liquid
 are visible only at the lowest energies, perhaps even only below the scope of the experimental data. In this case, the challenge is to identify 
 a framework which can account for such a rapid crossover into a classical regime, \rmn{where quantum mechanics mainly
 enters in the mode occupation numbers}. An alternative would be the absence
 of a qualitatively distinct low-frequency quantum spin liquid regime altogether. This could either happen intrinsically, if the 
 emergent low-energy description is amenable to a semi-classical description; or extrinsically, in that the quantum spin liquid
 behavior is so fragile in {practice} that disorder or coupling to phononic degrees of freedom destroys it entirely. There are many tantalizing open questions.
The minimal next step for which experimental input would be most valuable would be to consider materials 
with other values of spin---ideally both $S=1/2$ and higher spin values---as well as extending the experimental window further down towards  the asymptotic low-frequency behavior, if possible in a sample including minimal
disorder. 

\sectionn{Acknowledgements} 
We thank John Chalker, Chris Laumann and Johannes Reuther for helpful discussions; and Collin Broholm
for collaboration on the experimental side of the project.
This work was supported through the Institute for
Quantum Matter at Johns Hopkins University, by the
U.S. Department of Energy, Division of Basic Energy
Sciences, Grant DE-FG02-08ER46544, and by the Deutsche 
Forschungsgemeinschaft via grant SFB 1143. 
HJC also thanks 
Florida State University for start-up funds. 
We gratefully acknowledge the Johns Hopkins Homewood High Performance Cluster (HHPC) 
and the Maryland Advanced Research Computing Center (MARCC), 
funded by the State of Maryland, for computing resources. { A portion of this research used resources at the Spallation Neutron Source, a DOE Office of Science User Facility operated by the Oak Ridge National Laboratory.} 

\appendix
\section{Correspondence between classical and quantum spin wave theories}
\label{Appendix:Correspondence}
We discuss linear spin wave theory from the classical and quantum perspectives. We begin 
with a classical ground state (or generally a local minimum of energy), so that every spin $\mathbf s_i$ 
is in a state of equilibrium. It is convenient to define a local frame with three mutually orthogonal unit vectors
$\mathbf u_i$, $\mathbf v_i$, and $\mathbf w_i$, where $\mathbf u_i \!=\! \mathbf s_i/S$ points along the equilibrium 
direction. Small deviations of $\mathbf s_i$ from its equilibrium position can be parametrized in terms of two 
coordinates $x_i$ and $y_i$ as follows: 
\begin{eqnarray}
\mathbf s_i &=& \sqrt{S^2 - S(x_i^2 + y_i^2)} \, \mathbf u_i + \sqrt{S} (x_i \mathbf v_i + y_i \mathbf w_i)
\nonumber\\
&\approx& \left(S-\frac{x_i^2+y_i^2}{2}\right)  \mathbf u_i + \sqrt{S} (x_i \mathbf v_i + y_i \mathbf w_i).
\end{eqnarray}

The dynamics of variables $x_i$ and $y_i$ is governed by the Lagrangian
\begin{equation}
L = \sum_i \frac{1}{2} \left(y_i \frac{d x_i}{dt} - x_i \frac{d y_i}{dt}\right) 
	- U,
\label{eq:lagrangian}
\end{equation}
where $U$ is the potential energy encoding the spin interactions. Upon expanding it to the second order in the deviations from equilibrium, we obtain
\begin{equation}
L = \frac{1}{2} z^T \Gamma \frac{d}{dt} z - \frac{1}{2} z^T \mathcal H z, 
\label{eq:L-matrix-form}
\end{equation}
where $\mathcal H$ is a symmetric matrix, $\Gamma$ is a skew-symmetric matrix and $z$ is a column vector:
\begin{equation}
z \equiv 
	\left(
		\begin{array}{c}
			x_1 \\ 
			y_1 \\
			\vdots\\
			x_N \\ 
			y_N
		\end{array}
	\right),
\quad 
\Gamma = 
	\left(
		\begin{array}{ccccc}
			0 & -1 & \ldots & 0 & 0\\
			1 & 0 & \ldots & 0 & 0\\
			\vdots & \vdots & \ddots & \vdots & \vdots \\
			0 & 0 & \ldots & 0 & -1\\
			0 & 0 & \ldots & 1 & 0
		\end{array}
	\right).
\end{equation}

\subsection{Classical approach}
In the classical approach, the momentum- and energy-dependent spin correlation is defined as
\begin{equation}
\begin{aligned}
\mathcal S_{\text{classical}}^{\mu \nu} (\mathbf q, \omega) 
= & \frac{1}{2\pi N} \sum_{i,j = 1}^{N} \int_{-\infty}^\infty dt \,e^{-i \mathbf q \cdot(\mathbf r_i - \mathbf r_j) + i \omega t} \\
& \qquad \qquad \qquad \times \langle s_i^\mu (t) s_j^\nu (0)\rangle.
\label{eq:classical-def}
\end{aligned}
\end{equation}
Here $i, \, j \!=\! 1 , \dots, N$ are indices labeling lattice sites and  $\mu, \, \nu \!=\! x,y,z$ are indices for Cartesian spin components. 

Tthe classical equations of motion for the deviations $z$ are given by the Lagrangian~(\ref{eq:L-matrix-form}), which correctly describe the precession of spins (\ref{eq:torque}),
\begin{equation}
\Gamma \frac{d}{dt} z = \mathcal H z.
\label{eq:classical-eom}
\end{equation}

The solution to this equation (\ref{eq:classical-eom}) 
\begin{equation}
z(t) = \sum_\alpha c_\alpha \psi_\alpha e^{-i \omega_\alpha t}
\end{equation}
is a superposition with amplitude $c_\alpha$ of the normal modes $\psi_\alpha$ of the eigenproblem
\begin{equation}
(i \omega_\alpha \Gamma + \mathcal H ) \psi_\alpha = 0.
\label{eq:eigen-equation}
\end{equation}
This eigenproblem has the following properties: the eigenvalues come in pairs of real numbers $\omega_{-\alpha} \!=\! -\omega_\alpha$ and the eigenvectors $\psi_{-\alpha}^\dagger \Gamma \psi_{-\alpha} \!=\! -\psi_\alpha^\dagger \Gamma \psi_\alpha$; we thus choose the orthonormalization
\begin{equation}
\psi_\beta^\dagger (- i \Gamma )\psi_\alpha = \text{sgn}(\omega_\alpha) \delta_{\alpha \beta}.
\label{eq:normalization}
\end{equation}
As a result,
\begin{equation}
\psi_\beta^\dagger \mathcal H \psi_\alpha =\text{sgn}(\omega_\alpha) \omega_\alpha  \delta_{\alpha \beta} =  |\omega_\alpha|  \delta_{\alpha \beta}.
\end{equation}
and the potential energy is diagonalized in $c_\alpha$ 
\begin{equation}
U = \frac{1}{2} \sum_\alpha \omega_\alpha c_\alpha^\star c_\alpha.
\end{equation}
The Boltzmann distribution gives the thermal average of the amplitudes 
$\langle c^\star_\beta c_\alpha \rangle \!=\!  \delta_{\alpha \beta} / \beta |\omega_\alpha|$,
thus
\begin{equation}
\langle z_k (t) z_l (0) \rangle =
 \sum_\alpha \frac{[\psi_\alpha]_k \,[\psi_\alpha^\dagger]_l}{\beta |\omega_\alpha|} e^{-i \omega_\alpha t}, 
\end{equation}
where $k, \, l \!=\! 1, \dots, 2N$. Denote $\psi_\alpha^i \equiv ([\psi_\alpha]_{2i}, [\psi_\alpha]_{2i+1})^T$ and $\eta^\mu_i \equiv (v_i^\mu, w_i^\mu)$, 
The spin correlation function (\ref{eq:classical-def}) can be expressed as
\begin{equation}
\begin{aligned}
&\mathcal S_{\text{classical}}^{\mu \nu} (\mathbf q, \omega) =  \frac{S}{N} \sum_{i,j = 1}^{N}  \,e^{-i \mathbf q \cdot(\mathbf r_i - \mathbf r_j) } \\
& \qquad   \qquad \qquad   \times \sum_\alpha \frac{ (\eta^\mu_i  \ldotp  \psi_\alpha^i) \, (\eta^\mu_j  \ldotp  \psi_\alpha^j)^\star}{\beta |\omega_\alpha|} \delta (\omega-\omega_\alpha).
\label{eq:classical-correlation}
\end{aligned}
\end{equation}

\subsection{Quantum statistics}
\label{Appendix:quantum}
In the quantum approach, the spin correlation can be computed as the imaginary part of the retarded response function
\begin{equation}
\begin{aligned}
\mathcal{G}_+^{\mu\nu}  =\frac{1}{N} \sum_{i,j = 1}^{N} \int_{0}^\infty dt \,e^{-i \mathbf q \cdot(\mathbf r_i - \mathbf r_j) + i \omega t} (-i) \langle [s_i^\mu (t), s_j^\nu (0)]\rangle.
\label{eq:quantum-propagator}
\end{aligned}
\end{equation}
As the temperature goes to zero, 
\begin{equation}
\mathcal S_{\text{quantum}}^{\mu \nu} (\mathbf q, \omega) 
= \frac{1}{\pi} \lim_{T \rightarrow 0}(1- e^{- \beta \omega})^{-1}  \text{Im} \mathcal{G}_+^{\mu\nu} =  \frac{1}{\pi} \text{Im} \mathcal{G}_+^{\mu\nu}.
\label{eq:quantum-def}
\end{equation}

To evaluate the functional average over the ensemble, we take $t \rightarrow -i \tau$ in (\ref{eq:L-matrix-form}) to get the imaginary-time partition function
\begin{equation}
\mathcal Z = \int \mathcal D z \; \exp \left[ {- \frac{1}{2} \int_0^\beta d \tau \; z^T (-i \Gamma \frac{d}{d\tau} +\mathcal H) z} \right].
\label{eq:quantum-partition}
\end{equation}
In the representation of Matsubara frequencies $\omega_n \!=\! 2 \pi n/\beta$ ($n \in \mathbb{Z}$),
\begin{equation}
z (\tau) = \frac{1}{\sqrt{\beta}} \sum_{\omega_n} \zeta_n e^{-i \omega_n \tau}.
\end{equation}
We diagonalize the action in the partition function (\ref{eq:quantum-partition}) by decomposing $\zeta_n$ into orthonormal vectors $\psi_\alpha$ (\ref{eq:normalization}) and calculate the matrix elements of the propagator:
\begin{equation}
\langle \zeta_n \zeta_{n}^\dagger \rangle_{kl} = \sum_{\alpha} \frac{[\psi_\alpha]_k \, [\psi_\alpha^\dagger]_l}{-i \omega_n  + \omega_\alpha} \text{sgn}(\omega_\alpha).
\end{equation}
Replacing $i \omega_n$ by $\omega + i0^+$ gives the retarded response function
\begin{equation}
\langle z_k(\omega) z_l (-\omega) \rangle_+ = - \sum_{\alpha} \frac{[\psi_\alpha]_k \, [\psi_\alpha^\dagger]_l}{\omega - \omega_\alpha + i 0^+} \text{sgn}(\omega_\alpha).
\end{equation}
The spin correlation function (\ref{eq:quantum-def}) is thus 
\begin{equation}
\begin{aligned}
&\mathcal S_{\text{quantum}}^{\mu \nu} (\mathbf q, \omega)  = \frac{S}{N} \sum_{i, j = 1}^{N}  \, e^{-i \mathbf q \cdot(\mathbf r_i - \mathbf r_j) } \\
& \qquad \qquad \times  \sum_\alpha (\eta^\mu_i  \ldotp  \psi_\alpha^i) \, (\eta^\mu_j  \ldotp  \psi_\alpha^j)^\star\, \delta(\omega - \omega_\alpha)\text{sgn}(\omega_\alpha),
\label{eq:quantum-correlation}
\end{aligned}
\end{equation}
with the same notation as in (\ref{eq:classical-correlation}).
\subsection{Correspondence}
Comparing Eqs.~(\ref{eq:classical-correlation}) and (\ref{eq:quantum-correlation}), we arrive at a relation between the finite temperature classical calculation and the zero temperature quantum calculation for spin correlations under the linear spin wave framework, 
\begin{equation}
\beta \omega \mathcal S_{\text{classical}}^{\mu \nu} (\mathbf q, \omega) = \mathcal S_{\text{quantum}}^{\mu \nu} (\mathbf q, \omega),
\label{eq:correspondence}
\end{equation}
which is applicable at low temperature and positive energy transfer ($\beta \omega \gg 1$).

\section{Numerical methods}
In this work, we have studied the dynamical structure factor of the spin-1 pyrochlore material \ncnf{}
with a combination of classical Monte Carlo and molecular dynamics simulations (MD), stochastic "large-N" (SLN) and real-space linear 
spin wave theory (LSWT). Here we provide additional details needed to reproduce our data in the main text.

\subsection{Monte Carlo-molecular dynamics}
We study the dynamics of interacting classical spins by integrating the Landau-Lifshitz equation~\cite{LandauLifshitz}
\begin{equation}
	\frac{d}{dt} \mathbf{s}_i = -\mathbf{s}_i\times \frac{\partial H}{\partial \mathbf{s}_i},
\label{eq:torque}
\end{equation}
which describes the precession of the spin in the local exchange field. Here time $t$ is in the unit of meV$^{-1}$.

Following previous work~\cite{MoessnerChalker1998PRL,MoessnerChalker1998PRB,ConlonChalker2009}, we perform Monte Carlo plus molecular dynamics simulations 
with the Hamiltonian $H$ derived previously by some of us for \ncnf{}~\cite{PlumbBroholm2017} (see main text). The initial configuration (IC) of spins is 
drawn by a Monte Carlo (MC) run from the Boltzmann distribution $\exp(-\beta H)$ at temperature $T \!=\! 1.8~K$. Then, for each starting 
configuration the spins are deterministically evolved according to Eq.~(\ref{eq:torque}). This is done for many independent initial 
configurations and the result is averaged, 
\begin{eqnarray}
\langle s^{\mu}_i(t) s^{\nu}_j(0) \rangle =  \sum_{\text{IC from MC}} s^{\mu}_i(t) s^{\nu}_j(0) \big|_\text{IC}.
\label{eq:mc-md}
\end{eqnarray}

We simulated spins on the pyrochlore lattice with $N\!=\!16L^3$ sites, where $L^3$ is the number of cubic unit cells and we show the results for $L\!=\!8$. 
$6000$ initial configurations were used from independent Monte Carlo runs.

To compute the dynamical spin structure factor in practice, we evolve the Landau-Lifshitz equation (\ref{eq:torque}) for a long but finite time $T_s \!=\! 60~\mathrm{meV}^{-1}$ 
, using the fourth-order Runge Kutta method with discretized time steps. 
The time step $\delta t \!=\! 0.02$ meV$^{-1}$ is chosen to be large enough to allow us to reach large $T_s$ efficiently, yet small enough to 
ensure that the energy is constant during the entire time evolution (to an accuracy of roughly six digits in the energy per site). 

In order to compute energy-momentum spin correlations computationally efficiently, we perform a Fourier transform of the spin 
configurations during the time evolution 
\begin{equation}
	\tilde{s}^{\mu}({\bf q},\omega) = \frac{1}{\sqrt{N}}  \sum_{i} \frac{1}{T_s } \sum_t  e^{-i{\bf q} \cdot {\bf r}_j + i\omega t} s^{\mu}({\bf r}_i,t).
\end{equation}
Then (\ref{eq:classical-def}) is equivalent to 
\begin{eqnarray}
	\mathcal{S}^{\mu\nu}({\bf{q}},\omega) &=&  \frac{T_s}{2 \pi} \langle \tilde{s}^{\mu}({\bf q},\omega) {\tilde{s}^{\nu}}(-{\bf q},-\omega) \rangle.
\end{eqnarray}

As an overall check, we took the integration $\int d \omega \, \mathcal{S}^{\mu\nu}({\bf{q}},\omega)$ to compare with the static spin correlations 
$\mathcal{S}^{\mu\nu}({\bf{q}}) $ of the spin configurations sampled from Monte Carlo runs. They are consistent with each other.

\subsection{Stochastic model}
\begin{figure}[]
\includegraphics[width=\linewidth]{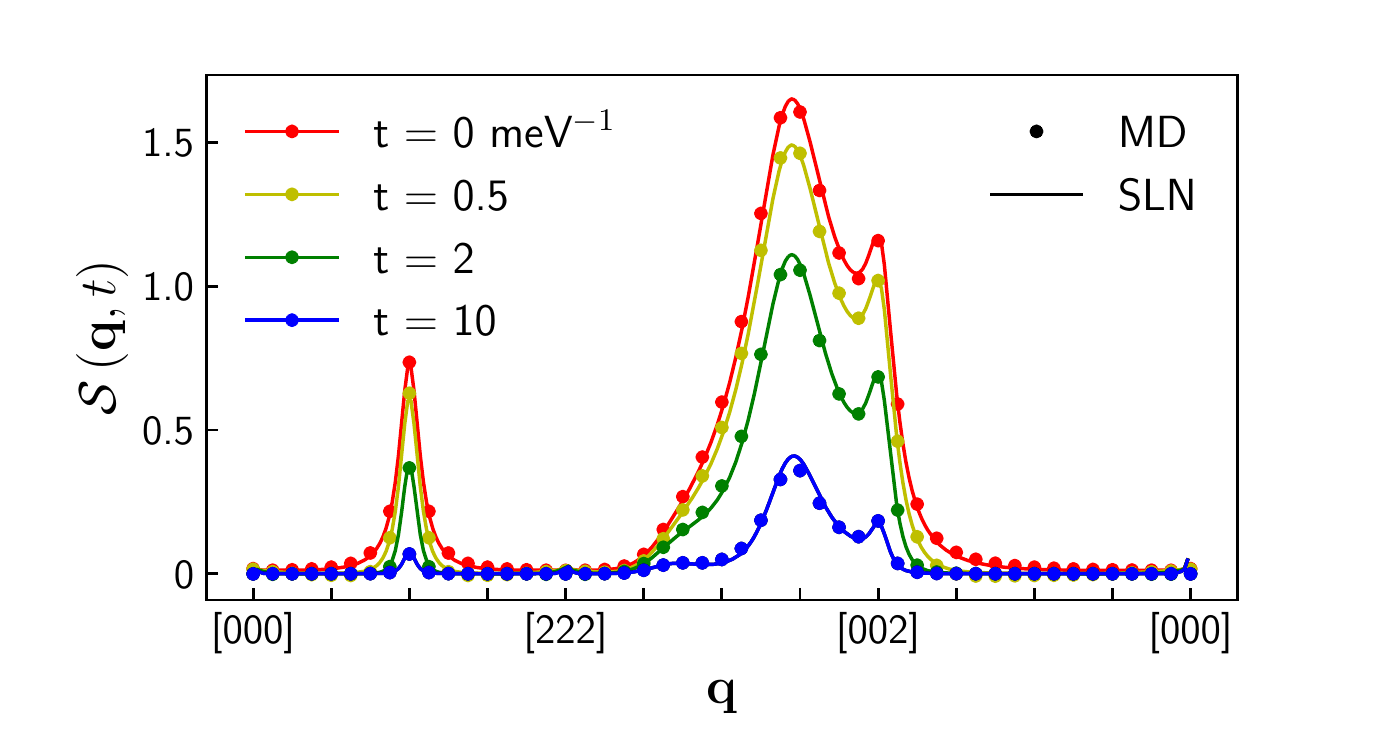}
\caption{(Color)
The time-dependent dynamical structure factor $\mathcal{S} (\mathbf{q},\omega)  \!= \! \sum_{\mu \nu} (\delta_{\mu\nu} - q_\mu q_\nu/q^2) \mathcal{S}^{\mu \nu} (\mathbf{q},\omega)$ from the MD (dots) and SLN (lines) along the path [000]~$\rightarrow$~[222]~$\rightarrow$~[220]~$\rightarrow$~[000] in momentum space at 4 different times including the static structure factor ($t  \!= \! 0$). This was used to fit the dynamical parameter $\gamma$ in the stochastic model, $\gamma  \!= \! 0.165$.}
\label{fig:fit-gamma}
\end{figure}

In the stochastic model, we study the hydrodynamic motion of the spins under the large-$N$ approximation ($N$ as of spin components). The spin configuration drifts under a generalized force, while a noise with Gaussian distribution provides thermal fluctuations~\cite{ConlonChalker2009}:
\begin{equation}
\frac{d}{dt} s^\mu_i = \gamma \sum_{j} \Delta_{ij} \frac{\partial E}{\partial s_j^\mu} + \xi^\mu_i(t),
\label{eq:stochastic}
\end{equation}
where $\gamma$ is a dynamical parameter to fit to the MD data (Fig.~\ref{fig:fit-gamma}) and $\Delta_{ij} \!=\! A^{(1)}_{ij} - z \delta_{ij}$ is the lattice Laplacian ($A^{(1)}_{ij}$ is the first-order adjacency matrix and $z$ is the coordination number).

The Boltzmann factor is
\begin{equation}
\beta E \equiv \sum_{ij} \sum_{\mu \nu} \frac{1}{2} ( \beta J^{\mu\nu}_{ij} + \lambda \delta_{ij} \delta^{\mu \nu}) s^\mu_i s^\nu_j ,
\end{equation} 
where $\lambda$ is the Lagrange multiplier fixing the average length of the "soft" spins.
As is the case in the self-consistent Gaussian approximation~\cite{GaraninCanals, PlumbBroholm2017},
$\lambda$ is solved self-consistently from
\begin{equation}{}
N S^2 = \sum_{\mathbf q \in \text{BZ}} \sum_\rho \frac{1}{\beta \epsilon_\rho (\mathbf q) + \lambda},
\end{equation}
where $\epsilon_\rho(\mathbf q)$ is the eigenvalues of the interaction matrix in the reciprocal space. 

The noise variables follow the independent Gaussian distribution $\langle \xi^\mu_i(t) \rangle \!=\! 0$ and $\langle \xi^\mu_i(t) \xi^\nu_j(t') \rangle \!=\! - (2 \gamma/\beta) \Delta_{ij} \delta^{\mu \nu} \delta (t-t')$, whose amplitude is determined by the fluctuation-dissipation theorem.{}

The lattice Fourier transform is performed on each of the four fcc sublattices ($a \!=\! 1, \dots, 4$),
\begin{equation}{}
\tilde{s}_a^\mu (\mathbf q) = \frac{1}{\sqrt{N_c}} \sum_{i \in a} e^{-i \mathbf q \cdot \mathbf r_{i}} s_{i}^\mu,
\end{equation}
where $N_c$ is the number of up- or down- tetrahedra and $4N_c  \!= \! N$ is the number of sites. The lattice Laplacian and the interaction matrix transform
accordingly into momentum space, denoted by $\Delta (\mathbf{q})$ and $J(\mathbf{q})$ respectively. Define 
$P \! \equiv \! - \Delta(\mathbf q) \otimes I_3$ and $Q  \! \equiv \! J(\mathbf q) + (\lambda/\beta) I_{12}$, where $I_n$ is an $n \times n$ identity matrix; 
the equation of motion in the momentum-energy space of the 12-component vector $\tilde{S} \! \equiv \! (\tilde{s}_1^x, \tilde{s}_1^y, \tilde{s}_1^z, \dots, \tilde{s}_4^x,\tilde{s}_4^y,\tilde{s}_4^z)^T$ can be expressed by matrix multiplication:
\begin{equation}
\tilde{S} (\mathbf q, \omega) = G(\mathbf q, \omega)  \, \tilde{\xi} (\mathbf q, \omega),
\end{equation}
where the Green's function is
\begin{equation}
G^{-1}(\mathbf q, \omega) = -i \omega I_{12} + \gamma PQ.
\end{equation}

Because $P$ and $Q$ are real and symmetric, $Q$ is positive definite and $P$ is semi-positive definite, there exists a similarity transformation under matrix $V$ to obtain a diagonal matrix $\Lambda$ with real and non-negative entries (generally $PQ \neq QP$),
\begin{equation}
Q P = V \Lambda V^{-1}.
\end{equation}
It can then be derived
\begin{equation}
\begin{aligned}
&\langle \tilde{S}_\alpha (\mathbf q, \omega)  \tilde{S}_\beta (-\mathbf q, -\omega)  \rangle  = \frac{2 \gamma}{\beta} (G P G^\dagger)_{\alpha \beta}\\
& \qquad \qquad = \frac{2 \gamma}{\beta} \left[ P V (\omega^2 + \gamma^2  \Lambda ^2 ) ^{-1} V^{-1} \right]_{\alpha \beta}.
\end{aligned}
\end{equation}
The classical spin correlations (\ref{eq:classical-def}) thus evaluate to be
\begin{equation}
\mathcal S^{\mu \nu}  (\mathbf q, \omega)  =
 \frac{1}{8 \pi} \sum_{\alpha \beta} \kappa^\mu_\alpha \kappa^\nu_\beta \langle \tilde{S}_\alpha (\mathbf q, \omega)  \tilde{S}_\beta (-\mathbf q, -\omega)  \rangle,
\label{eq:stochastic-correlation}
\end{equation}
where $\kappa^\mu \!=\! (1,1,1,1) \otimes \hat{\mathbf e}^\mu$ and the factor from $N \!=\! 4 N_c$ is accounted.

We further confirm the fluctuation-dissipation theorem is obeyed despite the subtlety from the non-commutation of the matrices. The static spin correlations upon integration of (\ref{eq:stochastic-correlation}) over energies is analytically the same as in our previous work~\cite{PlumbBroholm2017}.

\subsection{Real space linear spin wave theory}
\label{Appendix:rswt}
A typical application of linear spin wave theory is usually based on a classical 
ground state candidate that is derived or postulated. Often, the simplest ground states have a small unit cell of $n$ spins 
(say a pattern on a single tetrahedron, $n=4$, for the pyrochlore lattice~\cite{RossSavary}), which repeats in real space, so the wavevector is a good quantum number. The Hamiltonian in momentum space is block diagonalized and gives $n$ bands with sharp dispersions. However, the situation is markedly different for inhomogeneous ground states or states with very large unit cells. 
It becomes hard to track bands for large unit cells and for inhomogeneous states the loss of translational invariance means momentum is no longer a good quantum number.

This latter situation is typical for the classical Heisenberg model on the pyrochlore lattice. It has many ground states 
which satisfy the condition $\sum_{i \in \boxtimes}\mathbf{s}_i \!=\! 0$, and most of them are inhomogeneous. (Small anisotropic and further-neighbor interactions lift this degeneracy, but can still lead to the formation of many low-energy minima.) Thus here we consider finite lattice
clusters and perform linear spin wave theory in real space directly using the formalism given by Eq.~(\ref{eq:quantum-correlation}). A former general consideration can be found in Ref.~\cite{MaestroGingras}.

To closely mimic the situation in MD, we first assemble an ensemble of 
classical ground (or metastable low-energy) states for the \ncnf{} Hamiltonian by 
performing replica Monte Carlo runs ranging from very low temperature ($T\!=\!0.1$~K) 
to high temperature ($T\!=\!10$~K). This allows for good equilibration of spin configurations and largely prevents from the simulation 
getting "stuck". (Formally, if all independent runs are run for infinitely long they must all find the true ground state, this is not the case in practical finite runs.)
The last spin configuration encountered in each run at the lowest temperature is used as the starting configuration
for an iterative algorithm. This algorithm works by aligning one spin with its local exchange field; keeping all 
the other spins in the configuration fixed. One sweep consists of $N$ such moves (one for each spin). Multiple sweeps are performed 
until the spin directions stop changing completely, which guarantees that a stable low-energy minimum has been achieved.

\begin{figure}[bh!]
\includegraphics[width=0.9\linewidth]{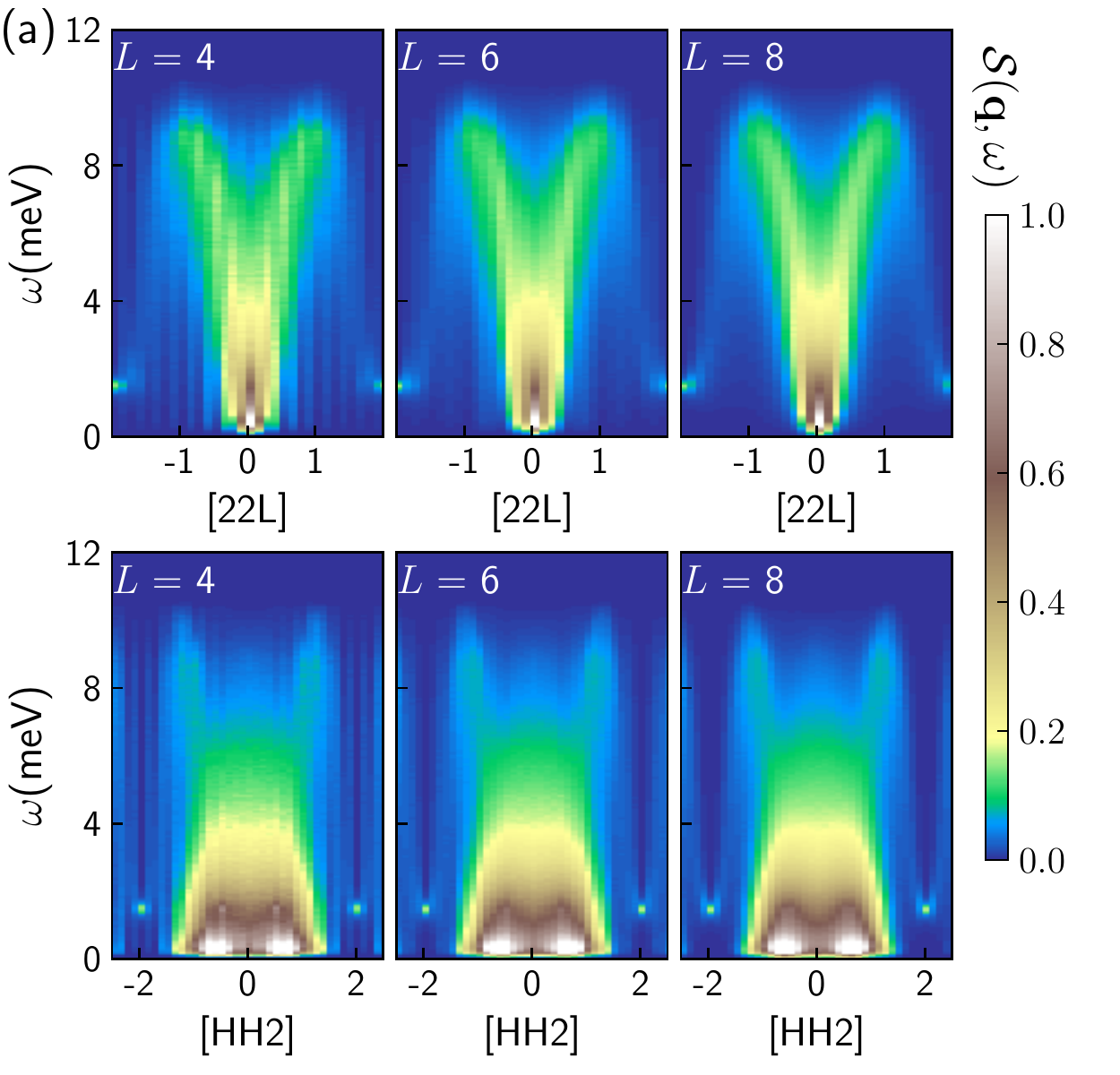}
\includegraphics[width=0.9\linewidth]{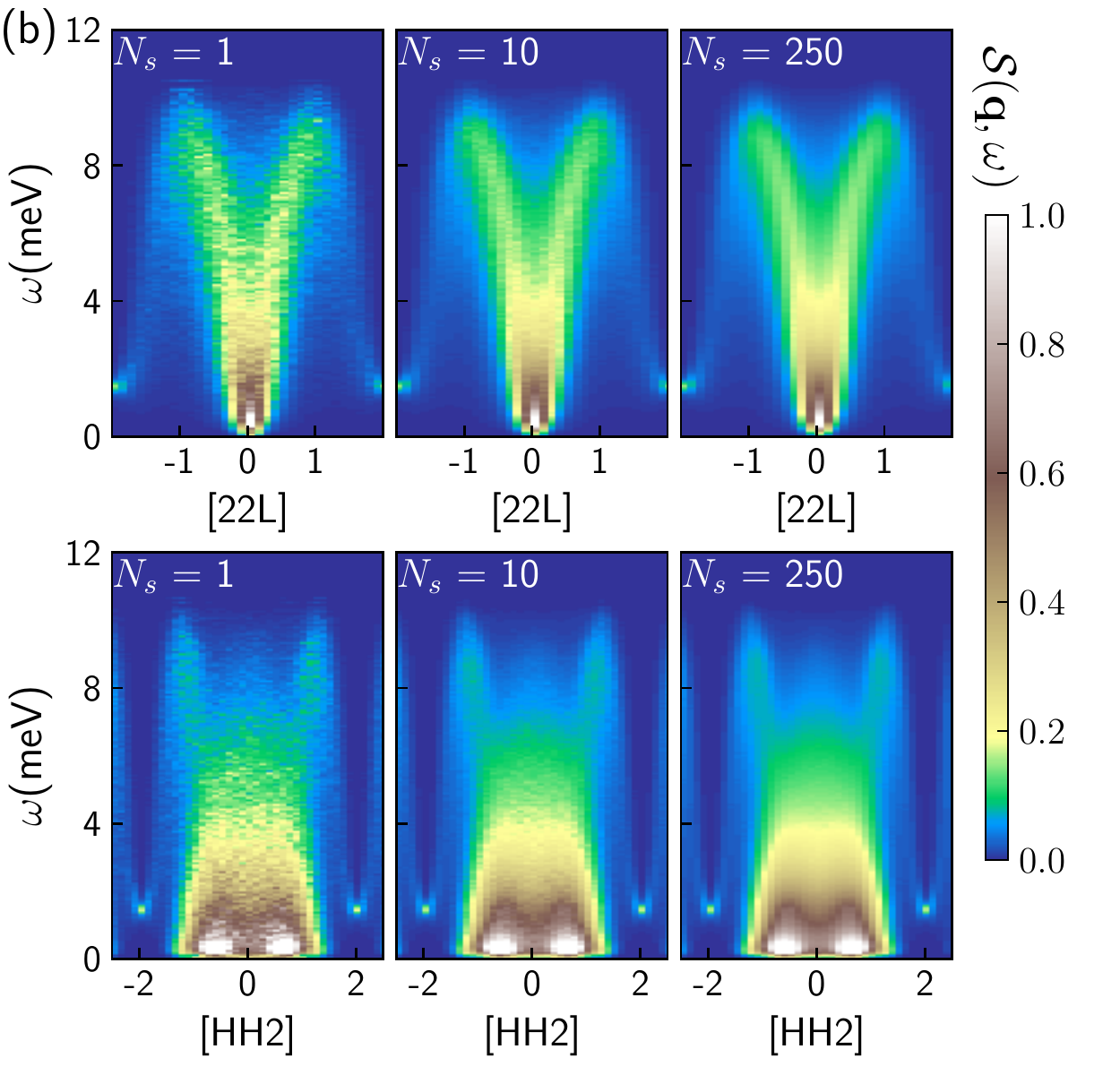}
\caption{(Color)
Real space linear spin wave theory results for different lattice sizes $N \!=\! 16L^3$ (a) and different numbers of samples $N_s$ with $L\!=\!8$ (b).}
\label{fig:size-sample}
\end{figure}

Then given a low-energy stable classical spin configuration, we construct and then directly diagonalize the potential $U$ in real-space. To evaluate the spin correlations (\ref{eq:quantum-correlation}) with discrete eigenvalues $\omega_\alpha$, we use a very narrow Gaussian function as an 
approximation to the Dirac delta function,
\begin{equation}
	\delta(\omega-\omega_\alpha) = \lim_{\epsilon \rightarrow 0} \frac{1}{\sqrt{2\pi} \epsilon} \exp \left[ { -\frac{(\omega-\omega_\alpha)^2}{2\epsilon^2} }\right],
\end{equation}
with $\epsilon=0.01$. 

Finally, we average the dynamical structure factor over $N_s$ low-energy spin configurations. The data in the main text is collected for $N_s \!=\! 250$ and $N \!=\! 16 L^3$ with $L\!=\!8$. 
Fig.~\ref{fig:size-sample}\textcolor{red}{(a)} shows our results for the dynamical structure factor along two different momentum cuts as a function of energy 
for $L\!=\! 4,6,8$. The finite size effects are small. Fig.~\ref{fig:size-sample}\textcolor{red}{(b)} shows the same cuts for different numbers of spin configurations used in the average, all with $L\!=\! 8$.

It can be shown that for a spin configuration with translational symmetry, our real-space LSWT is equivalent to the result in momentum-space, which presents sharp dispersions. Yet even for a single spin configuration~(see Fig.~\ref{fig:size-sample}\textcolor{red}{(b)}), the branches are broad. This confirms that the classical local minima found in the Monte Carlo simulation are indeed inhomogeneous.

We have also analyzed the inverse participation ratio (IPR) of the spin wave modes to understand the localization and delocalization effects. 
In analogy to the density distribution $\rho_e(\mathbf{r}) \!=\! | \psi_e(\mathbf{r})|^2$ of a given electron wavefunction,  which is normalized $\int d^3 \mathbf{r} \, \rho_e(\mathbf{r}) \!=\! 1$, we consider the normalization (\ref{eq:normalization}) and define the density distribution of the spin wave modes at energy $\omega_\alpha$ to be 
\begin{equation}
\rho_\alpha(\mathbf{r}_i) = (\psi_\alpha^i)^\dagger(-i \Gamma_2)\psi_\alpha^i,
\end{equation} 
where $\Gamma_2\!=\! \left( \begin{array}{cc} 0 & -1 \\  1 & 0 \end{array} \right)$ and  $\psi_\alpha^i \equiv ([\psi_\alpha]_{2i}, [\psi_\alpha]_{2i+1})^T$, corresponding to the two transverse spin deviations from the equilibrium direction. The IPR is given by
\begin{equation}
\text{IPR}(\omega_{\alpha}) \equiv \sum_{i = 1}^N |\rho_{\alpha}(\mathbf{r}_i)|^2.
\end{equation}

For a delocalized mode, we expect $\text{IPR}(\omega_{\alpha}) \sim 1/N$ while for a localized mode $\text{IPR} (\omega_{\alpha}) \sim O(1)$. 
In the main text Fig.~\ref{fig:ipr}, we show the IPR multiplied by the number of sites, i.e. $N \times \text{IPR}(\omega)$. The collapse of the values for different lattice sizes (except perhaps at the edges of the spectrum, especially the upper edge) indicates the delocalization of spin wave modes for a wide energy energy range. 

\section{Experimental data}
\label{Appendix:experiment}
The nonpolarized inelastic neutron scattering experiment probes the sum of the components of the spin correlation function that is perpendicular to the momentum transfer $\mathbf q$:
\begin{equation}
\mathcal {S}({\bf{q}},\omega) = \sum_{\mu,\nu} \left( \delta_{\mu\nu} - \frac{q_{\mu}q_{\nu}}{q^2} \right)\mathcal{S}^{\mu\nu}({\bf{q}},\omega).
\end{equation}
The measured intensity in the unit of scattering cross-section is given by 
\begin{equation}
	I({\bf q},\omega) = \left(\frac{\gamma r_o}{2}\right)^2|gf(q)|^2 \mathcal{S}({\bf q},\omega),
\end{equation}
where $f(|{\bf q}|)$ is the magnetic form factor, $g$ is the g-factor, and 
$\gamma r_0 = 0.539 \times 10^{-12} $~cm is the neutron magnetic scattering 
length. Throughout this work we use the dipole approximation to the Ni$^{2+}$ 
form-factor \cite{Brown:06} and an estimation of $g=2.28$~\cite{gfactorNi}.

Data presented in  Fig.~\ref{fig:patterns}\textcolor{red}{(b)} and Fig.~\ref{fig:dispersion} is 
identical to the published data set in Ref.~\cite{PlumbBroholm2017}. In addition to the previously published data set, we have also collected a new 
complementary data set with finer momentum and energy resolution. The inelastic 
neutron scattering data presented in Fig.~\ref{fig:patterns}\textcolor{red}{(a)} and 
Fig.~\ref{fig:methods-comparison} was collected on the cold neutron chopper 
spectrometer (CNCS) at Oak Ridge National Lab using fixed incident neutron 
energies of 2.5~meV, 6.59~meV, and 12~meV providing energy resolution (FWHM) at 
the elastic lines of 0.07~meV, 0.4~meV, and 0.95~meV respectively.  The same 
single crystal sample used for previous studies was mounted with the [HH0]
and [00L] directions in the horizontal scattering plane of the instrument.  
Data was collected with [00L] initially directed along the incident neutron 
beam and the sample rotated over 180$^{\circ}$. The sample was cooled to 350~mK
for all measurements. (No phase transition happens between 350~mK and 1.8~K and the dynamical structure factor at finite energy ($\gtrsim 0.1$~meV) is mostly temperature independent under the freezing temperature $T_f = 3.6$~K.) Data in  Fig.~\ref{fig:patterns}\textcolor{red}{(a)} has been symmetrized 
by folding about the [HH0] and [00L] axes. 

The data contains energy-dependent non-magnetic background intensity arising 
from incoherent nuclear scattering and scattering from the sample environment.  
This background contribution is estimated by taking a cut around the point 
${\mathbf q} = [0\,0\,3.65]$ and subtracted from the data in 
Fig.~\ref{fig:methods-comparison}. 

\bibliographystyle{apsrev4-1}
\bibliography{ncnf}

\end{document}